\documentclass[lettersize,journal]{IEEEtran}
\usepackage{amsmath,amsfonts}
\usepackage{amssymb}
\usepackage{latexsym}
\usepackage{cite}
\usepackage{amsmath,amssymb,amsfonts}
\usepackage{amsmath}
\usepackage{bm}
\usepackage{url}
\usepackage[pagewise, mathlines, switch]{lineno}
\usepackage{tabularx}
\usepackage{booktabs,rotating}
\usepackage{caption}
\usepackage{subcaption}
\usepackage{threeparttable}
\usepackage{multirow}
\newcolumntype{Z}{>{\raggedright\arraybackslash}X} 
\usepackage{algorithm}
\usepackage{algpseudocode}
\algrenewcommand\algorithmicrequire{\textbf{Input:}}
\algrenewcommand\algorithmicensure{\textbf{Output:}}
\usepackage[dvipsnames]{xcolor}
\usepackage{array}
\usepackage[caption=false,font=normalsize,labelfont=sf,textfont=sf]{subfig}
\usepackage{textcomp}
\usepackage{stfloats}
\usepackage{url}
\usepackage{verbatim}
\usepackage{graphicx}
\def\BibTeX{{\rm B\kern-.05em{\sc i\kern-.025em b}\kern-.08em
    T\kern-.1667em\lower.7ex\hbox{E}\kern-.125emX}}
\usepackage{balance}

\begin{document}
\title{Enhancing Super-Resolution Network Efficacy in CT Imaging: Cost-Effective Simulation of Training Data}
\author{Zeyu Tang, Xiaodan Xing and Guang Yang,~\IEEEmembership{Senior Member, IEEE}
\thanks{Zeyu Tang and Xiaodan Xing are co-first authors.}
\thanks{Z. Tang and X. Xing are with the Department of Bioengineering, Imperial College London, UK. G. Yang is with Department of Bioengineering and Imperial-X, Imperial College London, UK, Royal Brompton Hospital, London, and School of Biomedical Engineering \& Imaging Sciences, King's College London, UK. (send correspondence to x.xing@imperial.ac.uk and g.yang@imperial.ac.uk)}
\thanks{This study was supported in part by the ERC IMI (101005122), the H2020 (952172), the MRC (MC/PC/21013), the Royal Society (IEC/NSFC/211235), the Imperial College UROP, the NVIDIA Academic Hardware Grant Program, the SABER project supported by Boehringer Ingelheim Ltd, NIHR Imperial Biomedical Research Centre (RDA01), and the UKRI Future Leaders Fellowship (MR/V023799/1).}}
\maketitle

\begin{abstract}
 Deep learning-based Generative Models have the potential to convert low-resolution CT images into high-resolution counterparts without long acquisition times and increased radiation exposure in thin-slice CT imaging. However, procuring appropriate training data for these Super-Resolution (SR) models is challenging. Previous SR research has simulated thick-slice CT images from thin-slice CT images to create training pairs. However, these methods either rely on simplistic interpolation techniques that lack realism or sinogram reconstruction, which require the release of raw data and complex reconstruction algorithms. Thus, we introduce a simple yet realistic method to generate thick CT images from thin-slice CT images, facilitating the creation of training pairs for SR algorithms. The training pairs produced by our method closely resemble real data distributions (PSNR=49.74 vs. 40.66, p$<$0.05). A multivariate Cox regression analysis involving thick slice CT images with lung fibrosis revealed that only the radiomics features extracted using our method demonstrated a significant correlation with mortality (HR=1.19 and HR=1.14, p$<$0.005). This paper represents the first to identify and address the challenge of generating appropriate paired training data for Deep Learning-based CT SR models, which enhances the efficacy and applicability of SR models in real-world scenarios.

\end{abstract}


\section{Introduction}
\begin{figure}[ht]
    \centering
    \includegraphics[width=1.0\linewidth]{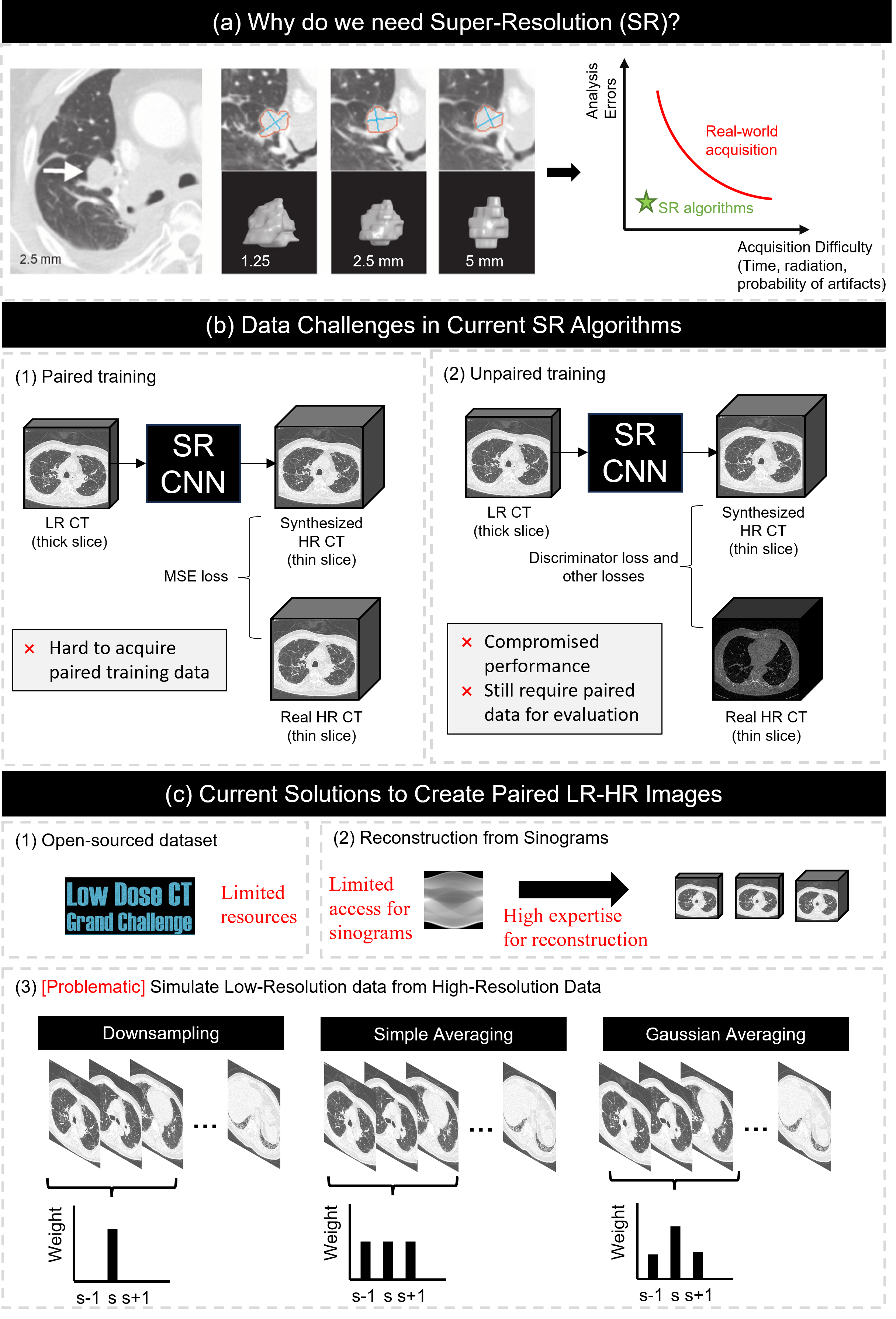}
    \caption{The image provides an overview of the necessity for our work. The rationale behind our work is to enhance the performance of CT super-resolution algorithms. (a) illustrates the need for SR in CT images; (b) highlights the difficulty in obtaining paired training data for SR CNNs, which can lead to compromised performance of the algorithms due to the lack of quality data for training and validation; (c) discusses the current methods used to simulate paired LR-HR images, pointing out the challenges and problems for failing to accurately represent the data.}
    \label{fig:motivation}
\end{figure}

Multi-slice helical CT has established itself as a formidable tool for the non-invasive and quantitative analysis of disease conditions, offering precise features crucial for evaluating patient conditions. The quality of quantitative analysis derived from CT images is highly contingent on the acquisition thickness, as exemplified in Figure \ref{fig:motivation}. Evidence from prior studies suggests that thin-slice (1.25 mm) CT scans have a superior capacity to distinguish benign from malignant solitary pulmonary nodules in comparison to thick-slice (5 mm) CT scans, demonstrating that thin slices are a richer source of information for radiomics analyses \cite{He2016,Zhao2014,Tan2012}. The level of detail provided by thin-slice imaging is instrumental for accurate diagnoses, enabling the detection of pathologies that may be overlooked with scans of lower resolution.

\textcolor{black}{However, balancing the clarity and detail of medical images with their file sizes presents a significant challenge. This balance impacts how much storage space is needed, how quickly files can be shared, and the amount of work required to analyze them. Despite the capability of contemporary CT scanners to reconstruct thin-slice images from the original sinogram data, the majority of images stored in the picture archiving and communication system (PACS) are not of thin-slice quality. This mainly stems from the need to manage file sizes, which helps reduce storage space, makes file sharing faster, and eases the workload for radiologists. However, for a wide range of clinical imaging tasks, doctors need high-resolution images to gain better insights, like the case we have shown above (Figure \ref{fig:motivation}a). Consequently, there is a gap between special clinical image applications and image quality in the PACS, highlighting the need for methods to enhance CT image resolution.}

One such innovation is the application of generative AI, or super-resolution (SR) algorithms in post-processing, which hold the promise of converting low-resolution (LR) thick-slice CT images into high-resolution (HR) thin-slice counterparts. Utilizing deep convolutional neural networks (CNN), researchers have devised models adept at modelling the complex nonlinear relationships between LR and HR data. The training of these deep SR models relies on paired LR-HR datasets for training to achieve optimal SR performance. Presently, the AAPM-Mayo's LDCT dataset emerges as the \textit{unique} public repository that provides such critical thin-thick slice pairs. In addition, since the scanning parameters are pre-defined, it is not normal to acquire both thin-slice and thick-slice images at the same time during a routine CT scan. 

\textcolor{black}{This scarcity of publicly available paired LR-HR training data, coupled with the limited practices in acquiring such paired datasets, forces researchers to generate simulated thick-slice CT images from their existing thin-slice CT datasets \cite{Panayides2020limitations, Li2021limitations}.} Those with access to high-resolution sinograms and their in-house CT reconstruction algorithms may reconstruct the sinograms with different reconstruction kernels, creating CT images with varying slice thicknesses \cite{Park2019}. Despite most scanners keeping their reconstruction methods proprietary and the specialized expertise required for reconstruction kernels, not all engineers have the privilege of accessing the original sinogram data. In many cases, researchers are limited to basic interpolation techniques, such as spline interpolation \cite{kudo2019} or slice removing \cite{luo2023autoencoder}, to produce realistic low-resolution images from high-resolution ones. Unfortunately, these simulated approaches frequently neglect key factors such as slice thickness and interval, leading to a misrepresentation of the distribution found in actual thick-slice images. Consequently, models trained on such data tend to perform poorly when applied to real-world thick-slice CT images.

This study sets out to introduce a novel simulation algorithm that can produce thick-slice CT images closely mirroring actual thick-slice scans, without the need for original sinograms or sophisticated reconstruction techniques. Our goal is to offer a nearly no-cost approach for generating paired LR-HR datasets that can support any CT super-resolution algorithm. As it stands, the AAPM-Mayo's LDCT dataset \cite{McCollough2017} is the only known public source of thin-thick slice pairs; hence, we used this dataset to validate our algorithm. We hypothesize that our simulation technique will generate thick-slice images that align more closely with real-life thick-slice images. Moreover, we propose that SR models trained with our simulated dataset will outperform those trained with other simulated datasets when applied to actual thick-slice images. The effectiveness of the SR models will be assessed using quantitative metrics including PSNR, MSE, SSIM and FID.

Our research is the first to tackle the challenge of paired training data in the CT super-resolution domain. In Section 2, we offer a thorough review of methods for simulating thick-slice images, and in Section 3, we introduce our proposed approach that incurs minimal costs to resolve this issue. Our efforts are dedicated not only to improving the applicability of CT super-resolution algorithms to real thick-slice CT images but also to deepening the understanding of the inherent non-linear relationship between LR and HR images.

\section{Related Works}
Deep learning has significantly transformed the landscape of medical imaging super-resolution, facilitating the development of models capable of learning intricate transformations from low-resolution to high-resolution images. Convolutional Neural Networks (CNNs) and Generative Adversarial Networks (GANs) are frequently employed frameworks in this context. It should be noted that our research does not concentrate on the methodologies of super-resolution techniques. We emphasize the generation of their paired training data. Therefore, this section will not provide a literature review on super-resolution methods but will outline prevalent simulation methods used to create LR-HR training pairs for super-resolution tasks.

\subsection{Conventional Simulation Methods}
\textbf{Direct Downsampling (Nearest Neighbor).} Unlike interpolation-based methods, direct downsampling reduces the number of slices by removing slices according to a certain ratio to emulate a thicker slice. This technique can include taking every second, third, or nth slice to represent an increased slice thickness. For example, Ge et al. \cite{Ge2023} simulated 1mm-3mm and 1mm-3mm CT image pair by performing direct downsampling on the thin-slice (1mm) data. Wang et al. \cite{Wang2021} generated thick slices by downsampling directly on the 1mm thin-slice images. Wu et al. \cite{Wu2022} simulated thick slices by downsampling directly on the 2.5mm thin-slice images. To further adjust the results, Mansoor et al. \cite{Mansoor2018} applied a 3D Gaussian smoothing filter on slices with 1mm thickness followed by downsampling to create slices with 4mm thickness on their in-house chest CT dataset. 

\textbf{Simple Averaging (Linear Interpolation).} In simple averaging, the pixel values of multiple contiguous thin slices are directly averaged to produce a single thick slice. While simpler to implement, this method assumes uniform contribution from each thin slice and can result in significant loss of detail, as no consideration is given to the specific intensity distribution or structural importance of each slice. For example, Park et al. \cite{Park2018} employed a method for thick-slice image simulation that involved averaging five slices with a 3mm thickness to create a single slice with a 15mm slice thickness on their in-house Brain CT dataset. In this approach, the middle slice was selected as the ground truth high-resolution image. Xie et al. \cite{Xie2021} employed a method for simulating thick-slice images by averaging three or seven slices with a 1mm thickness from their in-house brain CT dataset. This averaging process resulted in the creation of a single slice with a thickness of 3mm or 7mm, respectively.

\textbf{Gaussian Averaging (Gaussian Interpolation).} This technique involves the application of a Gaussian filter to the thin-slice images. The Gaussian filter gives more weight to the central slice and progressively less weight to the slices further away, based on the Gaussian distribution. This creates a smooth transition and can mimic the blurring effect seen in thicker slices. For example, Kudo et al. \cite{kudo2019} simulated various combinations of slice thickness and slice interval by reducing the number of slices to either 1/4 or 1/8. They applied spline interpolation and random Gaussian noise to the reduced slices.

\subsection{Limitations}
\label{sec:limitation}
\begin{figure}[!ht]
    \centering
    \includegraphics[width=1.0\linewidth]{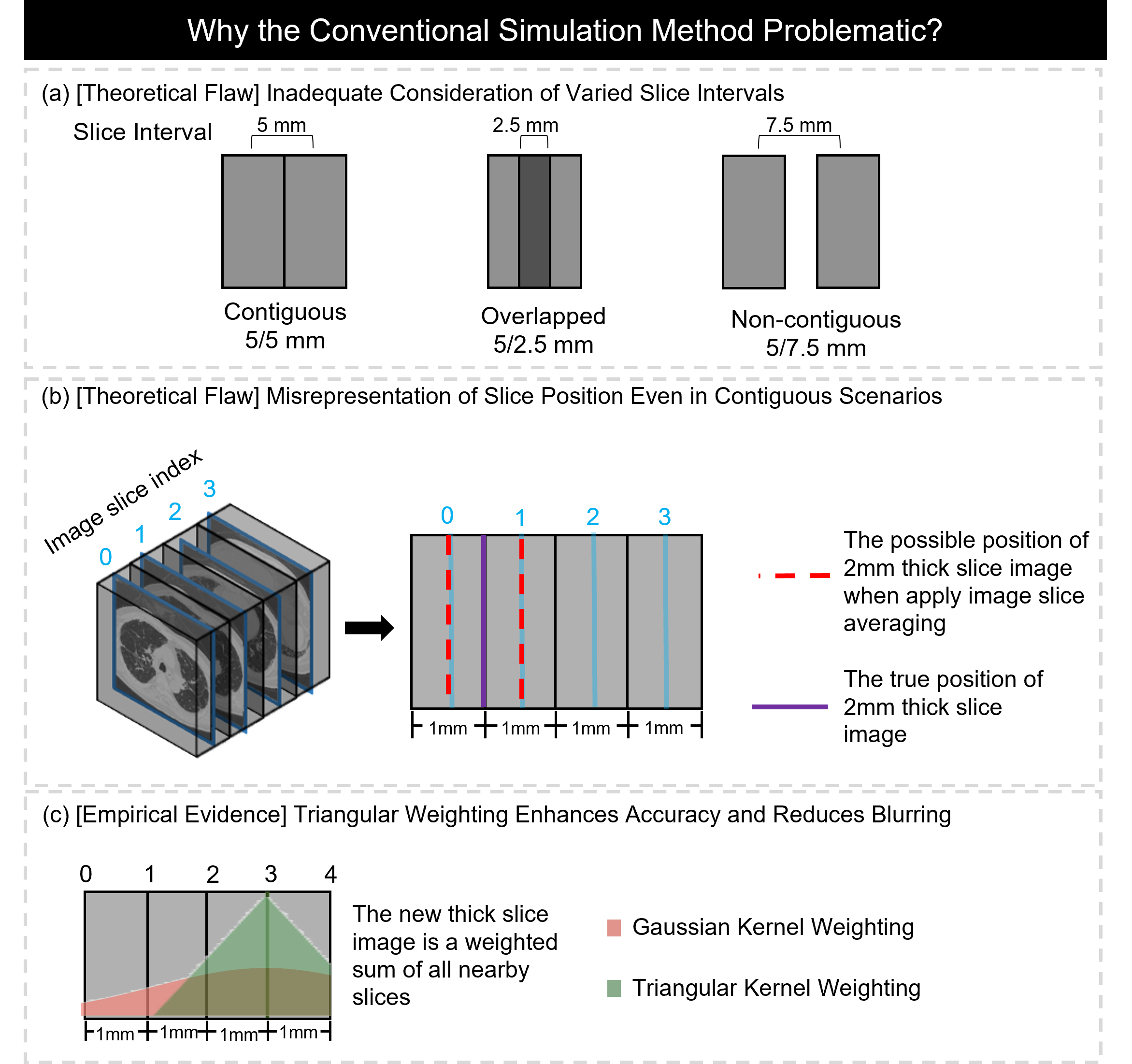}
    \caption{The limitations of conventional methods used for simulating thick-slice images from thin-slice CT scans. These traditional approaches can lead to inaccuracies in representing the true position (a,b) and quality (c) of the thick-slice images.}
    \label{fig:limitation}
\end{figure}
The underlying assumption of many conventional simulation algorithms is that the number of slices is directly proportional to the slice thickness. For instance, it is presumed that if an image with a 1mm slice thickness comprises slices, then to simulate a 3mm thick slice image, one would reduce the number of slices to $n/3$. However, this assumption is flawed when the scanning slices are not continuous, as is shown in Fig. \ref{fig:limitation} (a). An illustrative case is seen in the paired AAPM Mayo Clinic data for patient L286, where the images with 1mm slices total 525, but those at 3mm thickness only amount to 210 slices, not the expected 175 ($525/3$). This variance arises from overlapped scanning practices that result in a 2mm voxel spacing for 3mm thickness images. One major issue arising from this discrepancy in slice numbers is that if a model is trained to upscale the z-slices by a factor of, for example, three, it would never truly achieve an accurate reconstruction of the original slice distribution. This mismatch leads to a fundamental flaw in the model's ability to generalize from the training data to real-world scenarios.

Another theoretical shortcoming of conventional simulation algorithms is that they rely on image slice indexing, thereby disregarding the actual physical thickness of the images. For example, in Fig. \ref{fig:limitation} (b), when simulating 2mm thick images from 1mm thick images, both Gaussian Averaging and Direct Downsampling algorithms select an existing slice as the "centre slice". This results in misplacement of the thick slice's position, as shown by the blue line, instead of its true position in real 2mm thickness scans, indicated by the red line. This misplacement may lead to significant errors during the validation of SR performance on actual thick-slice datasets. 

Another limitation of conventional simulation algorithms is their dependence on image slice indexing, which ignores the actual physical thickness of the images. However, this approach misplaces the position of the thick slice, as demonstrated by the blue line. In reality, the correct position in actual 2mm thick scans is represented by the red line. This discrepancy can lead to significant errors when validating Super-Resolution (SR) performance on datasets with thicker slices.

In cases where voxel spacing and slice positioning are accurately established, the weighting kernel, which is designed to simulate the contribution of individual thin slices to the resultant thick slice in CT imaging, often remains empirically unverified. Traditional approaches like Gaussian distribution or spline interpolation are prevalent, yet our experimental data indicate that a triangular weighting strategy might yield superior results. 

\section{Methods}

\subsection{Position Correction}
The proposed method for simulating thick-slice CT images from their thin-slice counterparts starts from determining the new positions for the simulated thick slices, utilizing the slice interval $d$. Different from the traditional interpolation techniques mentioned in section \ref{sec:limitation}, we calculate the positions of the thick slices based on their actual physical locations according to The Patient-Based Coordinate System framework.

The spatial coordinates of each slice are documented in the DICOM header of the image. For a thin slice CT image, we first identify the coordinates of the starting $s$ and ending $e$ slices of the CT image volume. Then, the positions for the real thick slices are progressively calculated by adding increments of $d$, thus spanning the entire distance from $s$ to $e$. This approach ensures that the simulated thick-slice images are correctly placed in the physical space. The detailed procedure is outlined in Algorithm \ref{alg:slice_loc}.


\begin{algorithm}[!hbt]
\caption{\textcolor{black}{Determine Slice Locations}}\label{alg:slice_loc}
\begin{algorithmic}[1]
\Require Start position $s$, End position $e$, Interval distance $d$
\Ensure Array of slice locations $L_{\text{thick}}$
\State Initialize $L_{\text{thick}}$ as an empty list
\State Set initial point $p \gets s$
\While{ $s \leq p \leq e$}
    \State Add $p$ to $L_{\text{thick}}$
    \State Update point $p \gets p +  d$
    \EndWhile
\end{algorithmic}
\end{algorithm}

\subsection{Weighted Average}
Following the determination of the axial coordinates for all thick slices with $p$ representing the z-coordinate of each thick slice, the next step is to compute the contribution from each thin slice to the reconstruction of a thick slice. 

Mathematically, for each thick slice, we would like to compute the contribution of each thin slice, located at $l \in L_{\text{thin}}$, to the reconstruction of this thick slice whose position is noted as  $p \in L_{\text{thick}}$.

Drawing inspiration from the Weighted Filter Back Projection (wFBP) \cite{wFBP}, we have crafted a weighted sum algorithm designed to simulate thick-slice images from thin-slice counterparts. This is achieved through a triangular weighting function $g(p,l,t)$, with $t$ indicating the slice thickness. It should be noted that slice thickness $t$ is different from the slice interval $d$, as illustrated in Fig. \ref{fig:limitation}, and we use the slice thickness to compute the contribution to better simulate the contribution from each sinogram slice to the image slice. 

The function is mathematically defined as:
\begin{equation}
    g(p,l,t) = \max\left(0, 1 - \frac{{\left| p - l \right|}}{t}\right).
\end{equation}

The generated thick slices are weighted sums of the thin slices, normalized by the total weight of the slices used. The detailed steps for the generation of thick-slice images are illustrated in Algorithm \ref{alg:weighted_sum}. 

\begin{algorithm}[!hbt]
\caption{\textcolor{black}{Weighted Sum of Images for Thick Slices}}\label{alg:weighted_sum}
\begin{algorithmic}[1]
\Require Thick-slice locations $L_{\text{thick}}$, Thin-slice images $\bm{I}_{\text{thin}} \in \mathbb{R}^{512\times512\times d_{thin}}$, Slice thickness $t$
\Ensure Weighted sum images $\bm{I}_{\text{thick}}$
\State Initialize $\bm{I}_{\text{thick}}$ with a zero matrix $\mathit{O}\in \mathbb{R}^{512\times512\times d_{thick}}$
\For{each location $l$ in $L_{\text{thick}}$}
    \State Initialize total weight $w_{\text{total}}$ to 0
    \State Initialize weighted sum image $\tilde{i}$ to $\mathit{O}\in \mathbb{R}^{512\times512}$
    \For{each image $\bm{i}$ at location $p$ in $\bm{I}_{\text{thin}}$}
        \State $w_{\bm{i}} \gets g(p,l,t)$
        \If{$w_{\bm{i}} = 0$} \State Skip this iteration \EndIf
        \State Accumulate weighted image: $\tilde{i} \gets \tilde{i} + \bm{i} \cdot w_{\bm{i}}$
        \State Update total weight: $w_{\text{total}} \gets w_{\text{total}} + w_{\bm{i}}$
    \EndFor
    \State Normalize $\tilde{i}$ by total weight: $\tilde{i} \gets \frac{\tilde{i}}{w_{\text{total}}}$
    \State Append normalized $\tilde{i}$ to $\bm{I}_{\text{thick}}$
\EndFor
\State \textbf{return} $\bm{I}_{\text{thick}}$
\end{algorithmic}
\end{algorithm}

\begin{table*}[htbp]
\footnotesize
\setlength{\tabcolsep}{10pt}
\centering
\caption{Key data acquisition parameters for each exam type in TCIA LDCT-and-Projection-data}
\label{tab:dataset1}
\begin{tabularx}{\linewidth}{ccXXX}
\toprule
Scanner
& Parameters &  Head CT (N)  & Chest CT (C) & Abdomen CT (L)\\
\midrule
\multirow{5}{*}{\parbox{3cm}{GE Healthcare\\(Discovery CT750i)}} 
& Tube Potential (kV) & 120 & 80-120 & 80-120 \\
& Contrast Enhanced & No & No & Yes \\
& Field of View & 200-260 & 282-423 & 315-500\\
& Reconstruction Algorithm & Standard & Standard & Standard\\ 
& Thickness/Increment (mm) & 5/5 & 1.25/1 & 5/3\\
\midrule
\multirow{5}{*}{\parbox{3cm}{Siemens Healthineers \\(SOMATOM \\Definition AS+, \\SOMATOM \\Definition Flash)}} 
& Tube Potential (kV) & 120 & 120 & 100-120 \\
& Contrast Enhanced & No & No & Yes\\
& Field of View & 250 & 300-350 & 300-350 \\
& Reconstruction Kernel & H40 & B50 & B30\\ 
& Thickness/Increment (mm) & 5/5 & 1.5/1 & 5/3\\
\bottomrule
\end{tabularx}
\end{table*}
\begin{table*}[!htbp]
\footnotesize
\setlength{\tabcolsep}{10pt}
\centering
\caption{Key data acquisition parameters for each exam type in 2016 Low Dose CT Grand Challenge}
\label{tab:dataset2}
\begin{tabularx}{\linewidth}{ZcXX}
\toprule
Reconstruction Kernel
& Parameters  & Full Dosage (FD) & Quarter Dosage (QD)\\
\midrule
\multirow{4}{*}{\parbox{3cm}{D45}} 
& Tube Potential (kV)  & 100-120 & 100-120 \\
& Tube Current (mAs) & 200 & 50\\
& Contrast Enhanced & Yes & Yes\\
& Thickness/Increment (mm) & 1/0.8 and 3/2 & 1/0.8 and 3/2\\
\midrule
\multirow{4}{*}{\parbox{3cm}{B30}} 
& Tube Potential (kV) & 100-120 & 100-120 \\
& Tube Current (mAs) & 200 & 50\\
& Contrast Enhanced & Yes & Yes\\
& Thickness/Increment (mm) & 1/0.8 and 3/2 & 1/0.8 and 3/2\\
\bottomrule
\end{tabularx}
\end{table*}
\subsection{Super-Resolution (SR) Models}
\begin{figure}[!ht]
    \centering
    \includegraphics[width=1.0\linewidth]{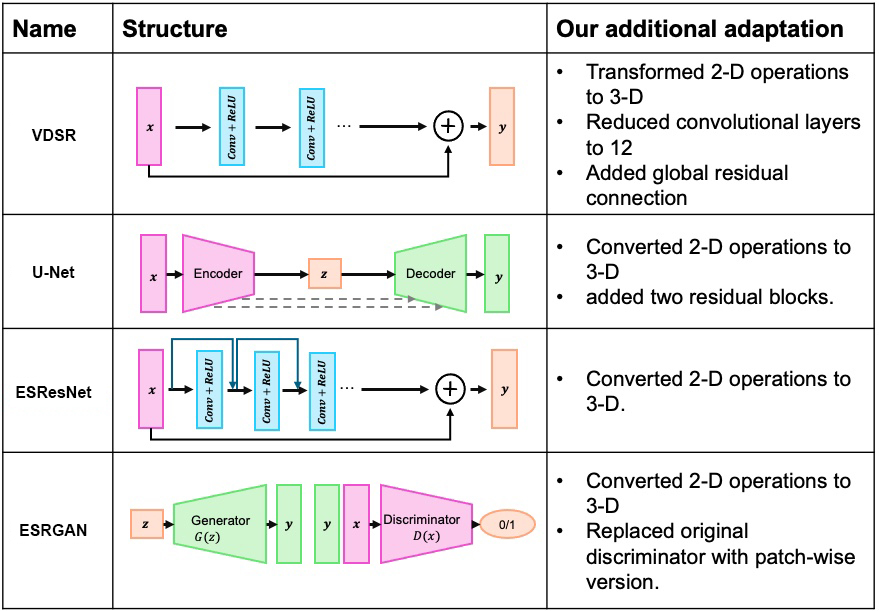}
    \caption{Illustration of our adaptations to four neural network architectures for super-resolution (VDSR \cite{VDSR}, U-Net \cite{UNet}, ESRresnet \cite{ESRGAN}, ESRGAN \cite{ESRGAN})}
    \label{fig:srmodel}
\end{figure}

To assess the quality of our simulated dataset for model training purposes, we chose four super-resolution models as benchmarks against our simulated thick-slice data. It is important to clarify that while the primary target of the study is not to introduce new SR architectures, we have re-engineered certain architectures to enhance the evaluation of SR performance as shown in Figure \ref{fig:srmodel}.
 
\begin{table*}[t]
\footnotesize
\centering
\begin{threeparttable}
    \setlength{\tabcolsep}{10pt}
    \caption{Comparison experiments on different thick-slice simulation methods}
    \label{tab:simulated_TS}
    \begin{tabular}{cccccc}
    \toprule
    {\textbf{Dataset}}
    & \textbf{Simulation Methods} &  \textbf{PSNR}  & \textbf{RMSE} & \textbf{SSIM} & \textbf{FID}\\
    \midrule
    2016 LDCT-GC (D45) & Simple Averaging & 40.6623 $\pm$ 3.1264\tnote{$\dagger$} & 0.0198 $\pm$ 0.0073\tnote{$\dagger$} & 0.9650 $\pm$ 0.0180\tnote{$\dagger$} & 0.0097 $\pm$ 0.0021\tnote{$\dagger$}\\
              & Gaussian Averaging & 30.0160 $\pm$ 5.8123\tnote{$\dagger$} & 0.0800 $\pm$ 0.0590\tnote{$\dagger$} & 0.8870 $\pm$ 0.0312\tnote{$\dagger$} & 0.1239 $\pm$ 0.0162\tnote{$\dagger$}\\
              & Direct Downsampling & 36.0757 $\pm$ 2.9024\tnote{$\dagger$} & 0.0334 $\pm$ 0.0131\tnote{$\dagger$} & 0.9489 $\pm$ 0.0204\tnote{$\dagger$} & 0.0243 $\pm$ 0.0050\tnote{$\dagger$}\\
              & Proposed & \textbf{49.7369 $\pm$ 2.5223} & \textbf{0.0068 $\pm$ 0.0020} & \textbf{0.9793 $\pm$ 0.0133} & \textbf{0.0074 $\pm$ 0.0034}\\
    \midrule
    2016 LDCT-GC (B30) & Simple Averaging & 42.0079  $\pm$ 2.8809\tnote{$\dagger$} & 0.0168  $\pm$ 0.0058\tnote{$\dagger$} & 0.9878  $\pm$ 0.0081\tnote{$\dagger$} & 0.0049 $\pm$ 0.0015\tnote{$\dagger$}\\
              & Gaussian Averaging & 33.4766 $\pm$ 5.9684\tnote{$\dagger$} & 0.0553 $\pm$ 0.0464\tnote{$\dagger$} & 0.9360 $\pm$ 0.0243\tnote{$\dagger$} & 0.0660 $\pm$ 0.0090\tnote{$\dagger$} \\
              & Direct Downsampling & 36.2342 $\pm$ 3.3025\tnote{$\dagger$} & 0.0334 $\pm$ 0.0150\tnote{$\dagger$} & 0.9711 $\pm$ 0.0119\tnote{$\dagger$} & 0.0164 $\pm$ 0.0018\tnote{$\dagger$}\\
              & Proposed & \textbf{48.5801 $\pm$ 7.3271} & \textbf{0.0108 $\pm$ 0.0099} &  \textbf{0.9945 $\pm$ 0.0059} & \textbf{0.0020 $\pm$ 0.0018}\\
              
    \bottomrule
    \end{tabular}
    \begin{tablenotes}
       \item [$\dagger$] represents statistical significance (with Wilcoxon signed-rank test p-value $< 0.05$) compared with the proposed method.
    \end{tablenotes}
\end{threeparttable}
\end{table*}
\begin{table*}[t]
\footnotesize
\centering
\begin{threeparttable}
    \setlength{\tabcolsep}{5pt}
    \caption{Comparison experiments on models trained by different simulated thick-slice images}
    \label{tab:sr_results}
    \begin{tabular}{ccccccc}
    \toprule
    {\textbf{Dataset}} 
    & \textbf{SR Model} & \textbf{Simulation Methods} &  \textbf{PSNR}  & \textbf{RMSE} & \textbf{SSIM} & \textbf{FID}\\
    \midrule
    2016 LDCT-GC & VDSR & Simple Averaging & 36.4722 $\pm$ 2.9112\tnote{$\dagger$} &  0.0320 $\pm$ 0.0131\tnote{$\dagger$} & 0.7617 $\pm$ 0.0649\tnote{$\dagger$} & 0.6344 $\pm$ 0.1071 \\ (D45) & & Proposed & \textbf{37.2176 $\pm$ 3.0876} & \textbf{0.0296 $\pm$ 0.0128} & \textbf{0.8140 $\pm$ 0.0640} & 0.6407 $\pm$ 0.0977 \\
    \cmidrule{3-7}
              & U-Net & Simple Averaging & 36.5079 $\pm$ 2.9129\tnote{$\dagger$} & 0.0319 $\pm$ 0.0133\tnote{$\dagger$} & 0.7683 $\pm$ 0.0642\tnote{$\dagger$} & 0.6655 $\pm$ 0.1140\\
              &  & Proposed & \textbf{37.9405 $\pm$ 2.1842} & \textbf{0.0262 $\pm$ 0.0071} & \textbf{0.8127 $\pm$ 0.0606} & 0.7072 $\pm$ 0.1755\\
    \cmidrule{3-7}
              & ESRResNet & Simple Averaging & 36.6322 $\pm$ 2.7383\tnote{$\dagger$} & 0.0312 $\pm$ 0.0120\tnote{$\dagger$} & 0.7859 $\pm$ 0.0650\tnote{$\dagger$} & 0.6407 $\pm$ 0.0659\tnote{$\dagger$}\\
              &  & Proposed & \textbf{37.8946 $\pm$ 2.5350} & \textbf{0.0267 $\pm$ 0.0094} & \textbf{0.8312 $\pm$ 0.0504} & \textbf{0.5556 $\pm$ 0.0701} \\
    \cmidrule{3-7}
              & ESRGAN & Simple Averaging & 36.8371 $\pm$ 2.4522\tnote{$\dagger$} & 0.0301 $\pm$ 0.0098\tnote{$\dagger$} & 0.7831 $\pm$ 0.0598\tnote{$\dagger$} & 0.6535 $\pm$ 0.1118\tnote{$\dagger$} \\
              &  & Proposed & \textbf{37.9786 $\pm$ 2.4597} & \textbf{0.0264 $\pm$ 0.0089} & \textbf{0.8271 $\pm$ 0.0477} & \textbf{0.5563 $\pm$ 0.0876}\\
    \midrule
    2016 LDCT-GC & VDSR & Simple Averaging & 37.2458 $\pm$ 4.6983\tnote{$\dagger$} & 0.0323 $\pm$ 0.0209\tnote{$\dagger$} & 0.8630 $\pm$ 0.0404\tnote{$\dagger$} & 0.3894 $\pm$ 0.0281\\ (B30) & & Proposed & \textbf{38.5657 $\pm$ 4.6613} & \textbf{0.0280 $\pm$ 0.0196} & \textbf{0.9060 $\pm$ 0.0315} & 0.3919 $\pm$ 0.0201\\
    \cmidrule{3-7}
              & U-Net & Simple Averaging & 37.4172 $\pm$ 4.4380\tnote{$\dagger$} & 0.0311 $\pm$ 0.0185\tnote{$\dagger$} & 0.8529 $\pm$ 0.0423\tnote{$\dagger$} & 0.3924 $\pm$ 0.0261\\
              & & Proposed & \textbf{39.7475 $\pm$ 3.5318} & \textbf{0.0226 $\pm$ 0.0112} & \textbf{0.9073 $\pm$ 0.0292} & 0.4238 $\pm$ 0.0767\\
    \cmidrule{3-7}
              & ESRResNet & Simple Averaging & 37.9217 $\pm$ 4.1180\tnote{$\dagger$} & 0.0289 $\pm$ 0.0171\tnote{$\dagger$} & 0.8788 $\pm$ 0.0379\tnote{$\dagger$} & 0.3258 $\pm$ 0.0374\tnote{$\dagger$}\\
              &  & Proposed & \textbf{40.2103 $\pm$ 3.9283} & \textbf{0.0223 $\pm$ 0.0150} & \textbf{0.9182 $\pm$ 0.0264} & \textbf{0.2608 $\pm$ 0.0220}\\
    \cmidrule{3-7}
              & ESRGAN & Simple Averaging & 38.2401 $\pm$ 3.6225\tnote{$\dagger$} & 0.0270 $\pm$ 0.0136\tnote{$\dagger$} & 0.8629 $\pm$ 0.0391\tnote{$\dagger$} & 0.3086 $\pm$ 0.0446\tnote{$\dagger$}\\
              &  & Proposed & \textbf{40.5083 $\pm$ 3.8736} & \textbf{0.0215 $\pm$ 0.0144} & \textbf{0.9199 $\pm$ 0.0278} & \textbf{0.2331 $\pm$ 0.0187}\\
    \bottomrule
    \end{tabular}
    \begin{tablenotes}
       \item [$\dagger$] represents statistical significance (with Wilcoxon signed-rank test p-value $< 0.05$) compared with the proposed method.
    \end{tablenotes}
\end{threeparttable}
\end{table*}

\section{Experimental Settings}
\subsection{Dataset}

This study principally utilized two data sets: the TCIA LDCT-and-Projection-data \cite{Moen2020} and the 2016 Low Dose CT Grand Challenge \cite{McCollough2017}. We simulated thick-slice data based on the TCIA LDCT-and-Projection data, employing it for the training phase. The 2016 Low Dose CT Grand Challenge was then deployed for testing purposes. Important features of these two datasets have been compiled and are presented in Table \ref{tab:dataset1}  and Table \ref{tab:dataset2} for reference.
\subsubsection{TCIA LDCT-and-Projection data}
This compilation includes 99 neuro scans (denoted by N), 100 chest scans (denoted by C), and 100 liver scans (denoted by L). Half of each scan category comes from a SOMATOM Definition Flash CT scanner, a product of Siemens Healthcare from Forchheim, Germany. The remaining scans, consisting of 49 for the head, 50 for the chest, and 50 for the liver, were captured with a Lightspeed Volume Computed Tomography (VCT) CT scanner from GE Healthcare, based in Waukesha, WI. Some data in this collection might be utilized to reconstruct a human face. In order to protect the privacy of individuals involved, those accessing the data are required to sign and submit a TCIA Restricted License Agreement upon usage.

\subsubsection{2016 Low Dose CT Grand Challenge (LDCT-GC)}
The dataset consists of 30 deidentified contrast-enhanced abdominal CT patient scans, which were obtained using a Siemens SOMATOM Flash scanner in the portal venous phase. The data comprises two types: Full Dose (FD) data and Quarter Dose (QD) data. Full Dose data corresponds to scans acquired at 120 kV and 200 quality reference mAs (QRM), while Quarter Dose data refers to simulated scans acquired at 120 kV and 50 QRM. The provided dataset includes various components: 1) Projection data for all 30 patient scans, including 10 cases for training purposes (both FD and QD) and 20 cases for testing (QD only). 2) DICOM images for the 10 training cases, encompassing FD and QD data, with reconstructions using 1 mm thick B30 and D45 kernels, as well as 3 mm thick B30 and D45 kernels. 3) DICOM images for the 20 testing cases, consisting of QD data only, with the same reconstruction configurations as the training cases (1 mm thick B30 and D45 kernels, and 3 mm thick B30 and D45 kernels).

\subsection{Evaluation Metrics}
\textcolor{black}{The Peak Signal-to-Noise Ratio (PSNR) and the Root Mean Square Error (RMSE) are two popular metrics used to measure the quality of an image, particularly for comparing the differences between the original and reconstructed data. PSNR is defined as the ratio between the maximum possible power of a signal and the power of corrupting noise that affects the fidelity of its representation. Mathematically, PSNR is calculated using the following formula:
\begin{equation}
    \text{PSNR} = 20 \cdot \log_{10}\left(\frac{\text{MAX}_I}{\sqrt{\text{MSE}}}\right),
\end{equation}
where $\text{MAX}_I$ is the maximum possible pixel value of the image. MSE stands for Mean Squared Error, which is the mean of the squared differences between the original and the reconstructed image. The mathematical expression is as follows:
\begin{equation}
\label{eq:mse}
    \text{MSE} = \frac{1}{mn} \sum_{i=0}^m \sum_{j=0}^{n} (I(i,j) - \hat{I}(i,j))^2,
\end{equation}
where $I(i,j)$ is the original value, $\hat{I}(i,j)$ is the predicted value, and $m \times n$ is the image dimension (i.e. total number of pixels in an image). RMSE is a quadratic scoring rule that measures the average magnitude of the error, and it is defined as the square root of the average squared differences between prediction and actual observation. The RMSE gives a relatively high weight to large errors, so it is especially useful when large errors are particularly undesirable. It can be calculated using equation \ref{eq:mse}:
\begin{equation}
    \text{RMSE} = \sqrt{\text{MSE}}.
\end{equation}
}
\begin{figure*}[!ht]
    \centering
    \includegraphics[width=1.0\linewidth]{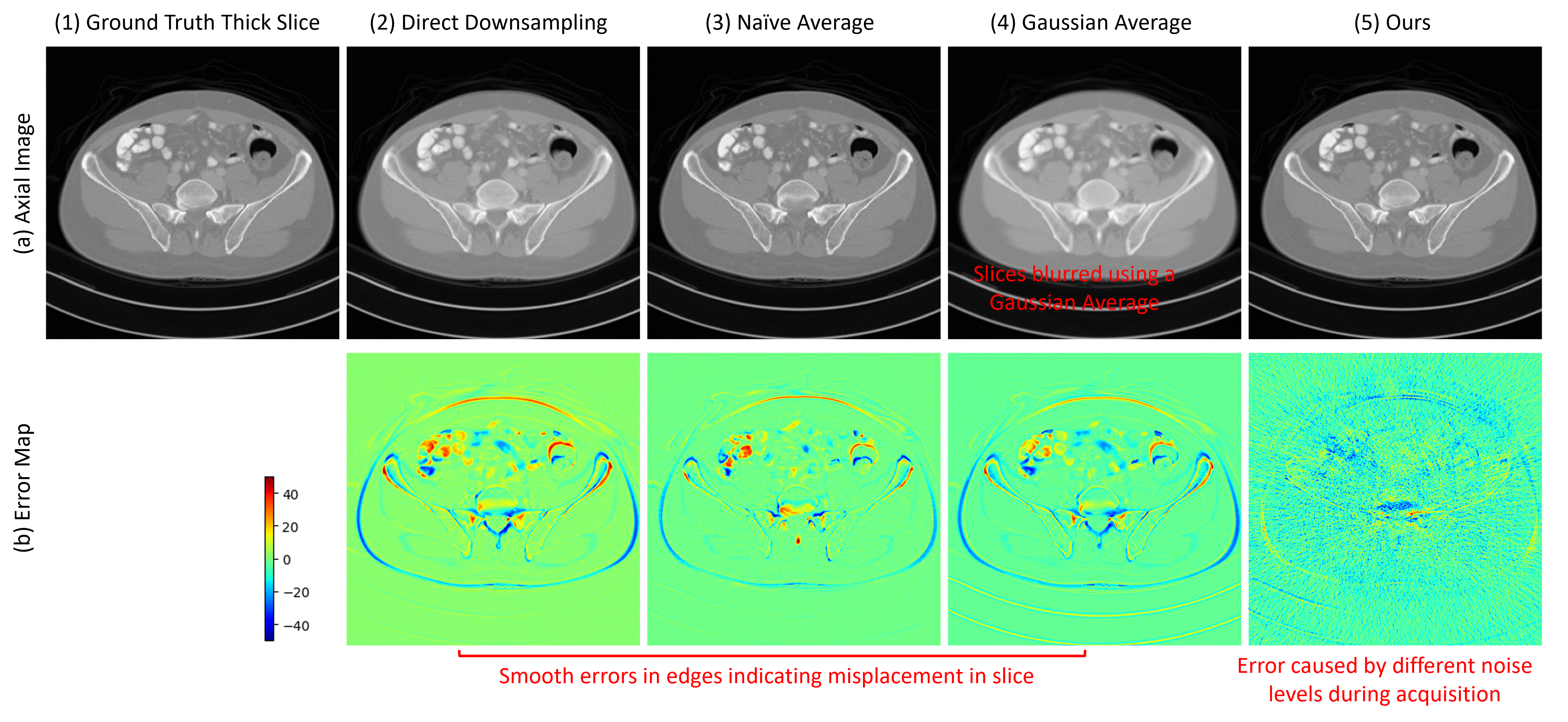}
    \caption{Example simulated thick slice using various simulation strategies and their corresponding error patterns. The Gaussian Average approach results in blurred imagery. The conventional simulations' error maps display smooth edge errors, indicating potential slice misalignment or displacement. In contrast, our method's error map demonstrates a markedly clearer outcome, with predominant noise patterns stemming from variations in slice thickness during acquisition and the Filtered Back Projection (FBP) process.}
    \label{fig:axial}
\end{figure*}

\subsection{Implementation Details}
The scripts used in this study were developed in Python3 and the PyTorch framework, and executed on Imperial College's HPC Clusters. Computations were run on an NVIDIA RTX 6000 GPU with 24 GB of memory. \textcolor{black}{Four distinct Super Resolution (SR) models: VDSR, U-Net, ESRResNet and ESRGAN (ESRResNet+PatchGAN Discriminator), were trained over 1000 epochs using the Adam solver (learning rate = 1e-4) with batch size 32 and augmented with random horizontal flips during training. All models were trained using L2 loss to minimize errors. For ESRGAN, the adversarial loss was added to the L2 loss.} The source code can be accessed at \url{https://github.com/ayanglab/Thick2Thin}.

\section{Results}

We assessed the performance of our proposed simulation method against Direct Downsampling, Simple Averaging, and Gaussian Averaging. The results from all conducted experiments, represented as mean $\pm$ standard deviation, are tabulated in this section. Unless otherwise specified, all results discussed here and in the following section are accomplished by simulating images with a thickness of 3mm from those of 1mm, utilizing the 2016 Low Dose CT Grand Challenge dataset. 

This section proves our hypothesis that our method generates the most realistic training pairs for synthetic images. This results in the most accurate SR performance of all SR CNNs when testing on real thick-slice images. 

\subsection{Image Fidelity Comparison}
The results presented in Table \ref{tab:simulated_TS} offer a detailed comparative analysis of various thick-slice simulation methods applied to two datasets from the 2016 Low Dose CT Grand Challenge, utilizing both the PSNR and the RMSE as the primary performance metrics.

The evidence strongly indicates that the proposed method outperforms Simple Averaging, Gaussian Averaging, and Direct Downsampling techniques in generating realistic thick slices from thin slices, applicable across both D45 and B30 reconstruction kernels. The validity of these superior results is supported by a Wilcoxon signed-rank test with a $p < 0.05$, signifying that the observed enhancements are statistically significant.

These findings corroborate our initial hypothesis, suggesting that the proposed simulation technique offers a more effective and accurate alternative for thick-slice simulations than conventional approaches. 

\subsection{Image Utility as Training Dataset}
We investigated our second hypothesis by training four different SR models using the data generated by top two simulation methods, in \ref{tab:simulated_TS}. According to table \ref{tab:simulated_TS}, we selected the Simple Averaging method for the utility comparison. After training the model on our simulated paired dataset, we tested these models' performance on true thick slice images to see if these models could reconstruct true thick slice images into thin slice images.

In each case, the performance of the SR models trained by the proposed simulation method outperforms those trained by the Simple Averaging simulation method, indicating improved image quality and lower error rates, respectively. The differences observed were statistically significant as determined by the Wilcoxon signed-rank test (p-value $<$ 0.05).

\section{Discussions}
\subsection{Visual Comparison of Fidelity}
In this section, we provide visual assessments of simulated thick slice images across axial planes as shown in Figure \ref{fig:axial}. We observed that traditional simulation methods consistently produce an error pattern, similar to motion-related artifacts. These error patterns are characterized by smooth, continuous errors along the image edges. Such patterns are not solely caused by slice misalignment but also by the averaging operations across slices, where the influence of adjacent slices is inaccurately estimated.

\begin{figure}[!ht]
    \centering
    \includegraphics[width=1.0\linewidth]{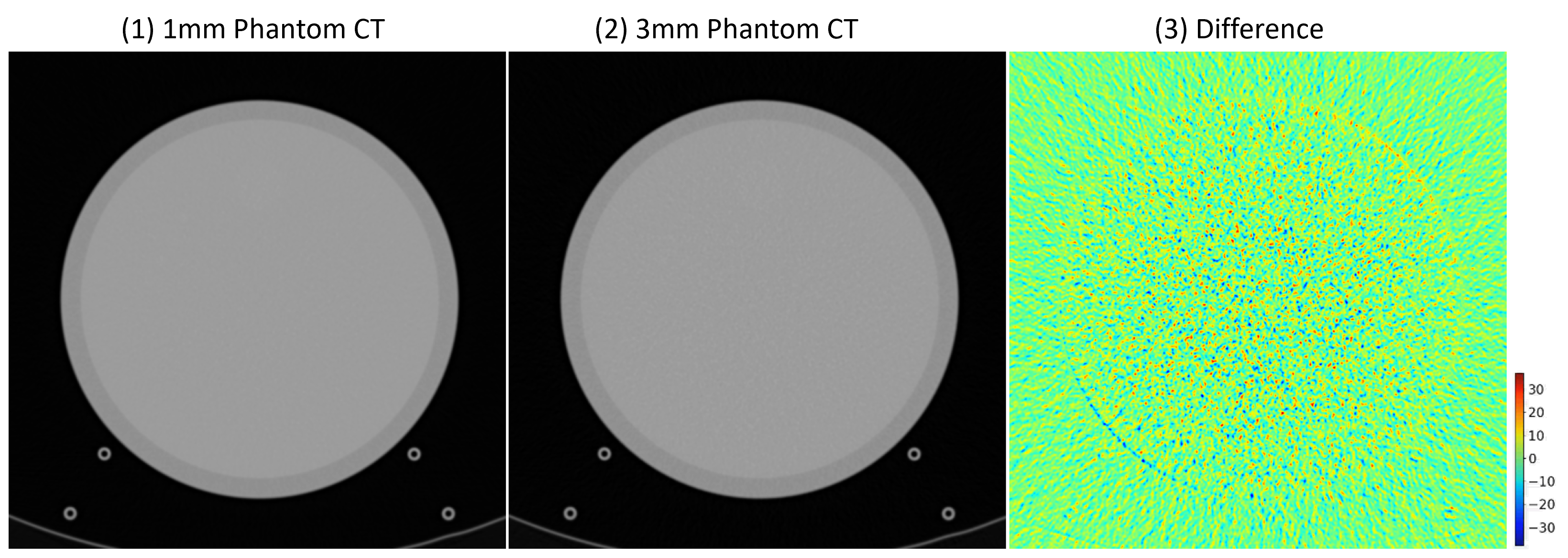}
    \caption{The full dose CT images reconstructed from B30 kernels acquired with different slice thickness and their differences.}
    \label{fig:noise}
\end{figure}

Conversely, the error map for our proposed method demonstrates a more accurate simulation, with minimal processing errors. Moreover, we noted that noise levels present during image acquisition and the back projection process could influence these minimal error patterns. Alshipli and Kabir's study \cite{alshipli2017effect} examined how slice thickness affects CT image noise, showing that thickness variations create distinct noise patterns. This is corroborated by the phantom images from the 2016 Low Dose CT Grand Challenge dataset, adhering to the protocol in \ref{tab:dataset2}, as depicted in Figure \ref{fig:noise}. These images display a comparable error pattern as in Fig. \ref{fig:noise}(3), suggesting that our algorithm's failure to differentiate between thin and thick slice images primarily leads to noise-related errors. However, the impact of this noise is minimal, leading to only slight discrepancies.

\subsection{Visual Comparison of Utility}
\begin{figure}[!ht]
    \centering
    \includegraphics[width=1.0\linewidth]{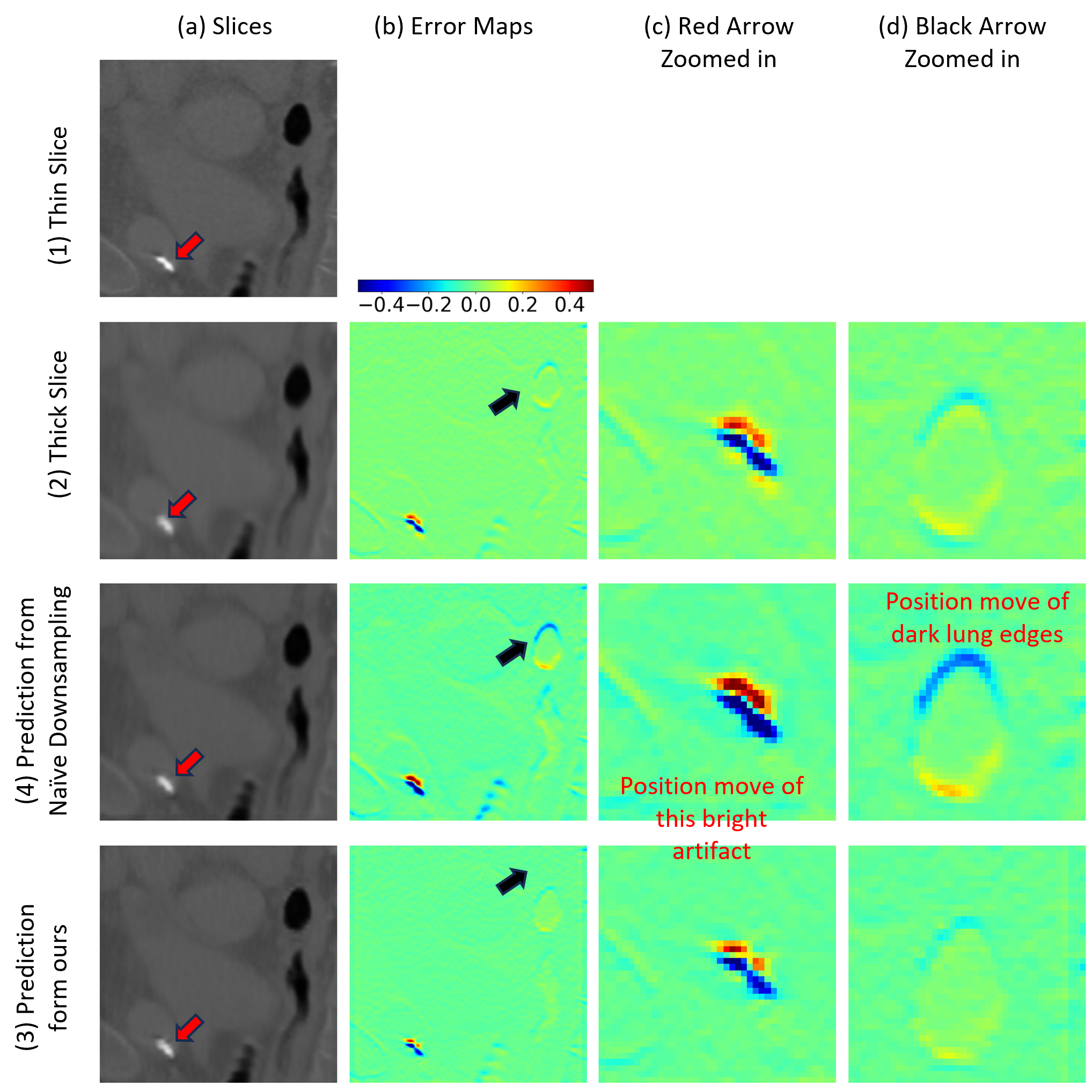}
    \caption{\textcolor{black}{Super resolved slices using the ERSGAN model, which was trained on simulated data generated by a conventional naive method (3) and ours (4). The images are compared with the original thin (1) and thick (2) slices. The arrows highlight areas where our method's superiority is evident}.}
    \label{fig:sr_performance}
\end{figure}

In this section, visual evaluations of super-resolved images are presented in Fig. \ref{fig:sr_performance}, which were generated by ERSGANs trained on data simulated using both our method and the conventional naive approach. We have chosen to feature ERSGAN due to its superior performance in comparison to other models. To match the resolution of the thin slice, the thick slice image has been trilinearly interpolated in the figures provided.

The red arrows highlight an image artefact, which has blurred edges in thick slices (2a). Interestingly, we note that in super-resolved images trained on data from the direct downsampling method, the position of this artefact has shifted significantly upward and to the right, as depicted by the red and black errors in (4b). This shift, along with the erroneous edge movement (black arrows), suggests that the incorrectly simulated thick slice data misdirect the training of super-resolution models, leading to an inaccurate mapping of the actual voxel locations during the super-resolution process. 

\begin{table*}[!t]

\footnotesize
\centering
\begin{threeparttable}[ht]
\centering
\setlength{\tabcolsep}{5pt}
\caption{Multivariate Cox Regression on radiomics features extracted from super-resolved images.}
\label{tab:survival}
\begin{tabular}{@{}lllll@{}}
\toprule
Covariate & \multicolumn{1}{c}{exp(coef)} & \multicolumn{1}{c}{exp(coef) Upper 95\%} & \multicolumn{1}{c}{p-value} \\ 
\midrule
\textbf{Thick} & & & \\
Elongation & 0 & 22.15 & 0.2 \\
Maximum2DDiameterColumn & 1.01 & 1.03 & 0.67 \\
Maximum2DDiameterSlice & 1.01 & 1.04 & 0.56 \\
MeshVolume & 1 & 1 & 0.82 \\
First order 10Percentile & 1.22 & 1.65 & 0.2 \\
\addlinespace
\textbf{Naive} & & & \\
Elongation & 0 & 3.57E+06 & 0.38 \\
Maximum2DDiameterColumn & 1.04 & 1.15 & 0.36 \\
Maximum2DDiameterSlice & 0.94 & 1 & 0.04 \\
MeshVolume & 1 & 1 & 0.82 \\
First order 10Percentile & 0.76 & 0.93 & 0.01 \\
\addlinespace
\textbf{Ours} & & & \\
Elongation & 4.05 & 2.52E+11 & 0.91 \\
Maximum2DDiameterColumn & 0.89 & 0.98 & 0.02 \\
\textbf{Maximum2DDiameterSlice} & \textbf{1.19} & \textbf{1.31} & $\mathbf{<0.005}$ \\
MeshVolume & 1 & 1 & 0.77 \\
\textbf{First order 10Percentile} & \textbf{1.14} & \textbf{1.24} &$ \mathbf{<0.005}$ \\
\bottomrule
\end{tabular}
        \begin{tablenotes}
           \item Concordance = 0.85
        \end{tablenotes}
\end{threeparttable}
\end{table*}

\subsection{Empirical Support on the Slice Correction}
\textcolor{black}{To clarify the rationale behind the slice correction step (Algorithm \ref{alg:slice_loc}), we performed an analysis to quantify the similarity between thick slice images and their corresponding thin slice images by calculating the mean squared error (MSE) between them. The objective was to assess how a single thick slice image represents the cumulative information from all nearby thin slice images. This evaluation was designed to approximate how contributions from individual thin-slice images aggregate to form a specific thick-slice image.}

\begin{figure}[!ht]
    \centering
    \includegraphics[width=1.0\linewidth]{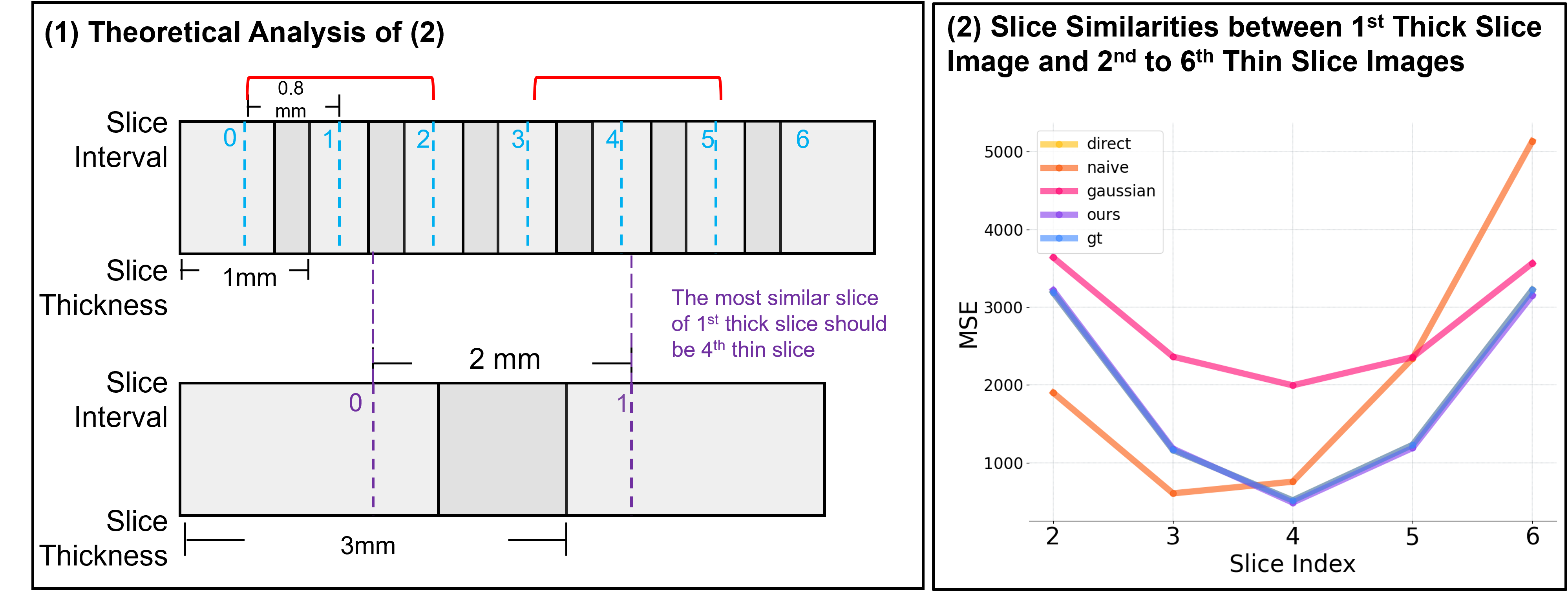}
    \caption{Slice similarity comparison between thin slices and a selected simulated thick slice. As is shown, our proposed method simulates thick slices with the most realistic contributions from thin slices.}
    \label{fig:similarity}
\end{figure}

\textcolor{black}{Within this framework, Figure \ref{fig:similarity} presents the analysis focused on the first axial slice (starting from the top of the lung) in a low-resolution computed tomography (CT) series. We calculated the MSE between simulated 3mm thick slices and axial slices 2 to 6 from corresponding 1mm thick CT images. Figure \ref{fig:similarity}(1) displays the axial slices for both the 1mm thin and 3mm thick slices, with slice spacings of 0.8mm and 1mm, respectively. }

\textcolor{black}{As is shown in Figure \ref{fig:similarity}(1), the axial thin slice most similar to the first thick slice should be slice number 4 when the spacing is considered. Contrarily, simpler methods like direct downsampling and naive downsampling, which do not the physical locations, inaccurately designate slice number 3 as the most similar, based on a flawed assumption that mistakenly applies a factor of 3 to simulate the transition from 1mm to 3mm slices.}

\textcolor{black}{As in Figure \ref{fig:similarity} (2), our method, incorporating a physical space location correction based on the precise physical coordinates from the DICOM header, demonstrates the most authentic pattern of similarity to the thin slice images.}

\subsection{Clinical Relevance}
The primary clinical benefit of super-resolved CT images lies in their ability to facilitate accurate quantification analyses of anatomical structures, typically necessitating HRCT. To illustrate this, we applied an airway segmentation model \cite{fuzzy_airway}, which was originally trained on HRCT, to both the original thick-slice CT images and super-resolved CT images produced by the ESRGAN model. Our goal was to evaluate the airway segmentation outcomes from these images to determine their efficacy in providing precise airway segmentation, thereby aiding the quantification analysis of thick-slice CT images.

The additional dataset used in the clinical relevance analysis is The Australian IPF repository (AIPFR), which has ethical approval from the Sydney Local Health District (protocol no. X14-0264). The diagnosis for each patient includes IPF, probable IPF, alternative diagnosis, and other fibrotic patterns based on the 2018 ATS/ERS/JRS/ALAT IPF guideline statement. We randomly selected 36 patients with an average slice thickness of 2.27 (STD=0.44) mm. The average age is 72.09 (STD=9.11).

The initial step involved applying ESRGAN models, trained on both naive and our simulation, to these thick-slice images, followed by airway segmentation inference. Subsequently, we filtered out radiomics features with low variance and selected five pertinent features relevant to mortality. The features we analysed include Elongation, the extent to which the shape of an ROI in the medical image is elongated; Maximum 2D Diameter Column, the largest possible diameter measured in the coronal plane of the airways; Maximum 2D Diameter Slice, the largest diameter measured in the axial plane of the airways; Mesh Volume, the volume of the airways based on a mesh approximation; and First order 10th Percentile, a statistical feature derived from the first-order intensity histogram of the airways. Multivariate analysis was then performed on these features, as presented in Table \ref{tab:survival}.

The analysis revealed that only the features extracted from the super-resolved images from the model trained on our simulation algorithm correlated with mortality. This finding suggests that our algorithm holds significant potential for retrospective studies involving datasets acquired exclusively in thick slices.

\subsection{Benchmarking four different models}
In our research, we conducted comparisons among various models to evaluate their SR effectiveness. The model that emerged as the most effective was the ESRGAN. This highlights the significant role of perceptual loss in SR models. For the benefit of other researchers, we have provided a summary of their efficiencies in Table \ref{tab:efficiency}, serving as a preliminary guide. The models will be open-sourced in \url{https://github.com/ayanglab/Thick2Thin}.

\renewcommand{\arraystretch}{1.5}
\begin{table}[]
\caption{The inference time, parameter size and performance of the 3D SR models implemented and refined by our paper. The performance is ranked by the mean value of the PSNR of these models on all testing datasets.}
\label{tab:efficiency}
\begin{tabular}{p{1.5cm}p{1.5cm}p{1.5cm}p{2cm}}
\toprule
\textbf{Model Name} & \textbf{Inference Time (s/image)} & \textbf{Parameter Size (MB)} & \textbf{Avg Performance Rank} \\\hline
VDSR & 9 & 298 & 4 \\\hline
U-Net & 6 & 4 & 3 \\\hline
ESRResNet & 8 & 40 & 2 \\\hline
ESRGAN & 8 & 83 & 1\\\hline
\end{tabular}
\end{table}

\subsection{Limitations and Future Works}
One of the significant limitations stems from the reliance on the AAPM-Mayo's LDCT dataset, which, to the best of our knowledge, is the only public resource providing thin-thick slice pairs. The performance of our proposed method has been evaluated primarily with respect to recreating the specific image characteristics of the AAPM-Mayo's LDCT dataset. It remains unclear how well our algorithm would perform with images obtained from patients with different health conditions. 


In this study, the chosen models are specifically selected for their stable performance and suitability for 3D medical imaging, owing to their minimal GPU memory requirements. However, it is important to note that other state-of-the-art 2D SR methods \cite{georgescu2023multimodal,chen2023activating}, not covered in this paper, also exist. We implemented a slice-wise Denoising Diffusion Probabilistic Model (DDPM) \cite{ho2020denoising} and attempted to enhance the resolution of 3D medical images on a 2D basis. Unfortunately, our findings indicate that the performance of this approach does not match that of comprehensive 3D volume methods. Thus, in the future, we will further investigate other novel SR algorithms and evaluate their applicability to 3D medical images.

Finally, we assessed the clinical capability of super-resolved CT images through inferencing segmentation models that outline anatomical structures, including the airway tree \cite{fuzzy_airway}. However, the lack of detailed annotations for these images posed a challenge, preventing us from using conventional metrics like the DICE coefficient for quantitative evaluation. Moving forward, we plan to extend our evaluation to additional clinical tasks using our method to comprehensively analyse the utility \cite{xing2023you} and accuracy of our algorithm once detailed annotation is obtained. This will involve a comprehensive exploration of the capabilities of SR models in clinical trial settings.

\section{Conclusions}
Our research introduces a simulation algorithm that improves the applicability of super-resolution algorithms on real thick-slice CT images with minimal computational overhead. The algorithm demonstrates the ability to create simulations that closely match actual thick-slice CT images, providing deep SR models with high-quality training data that yield superior results on genuine thick slices.

The key innovation in our study is a slice correction module that addresses the misalignment commonly found in traditional simulation approaches, allowing for the generation of precise training pairs for SR models. Another innovation, drawing on the principles of Weighted Filtered Backprojection (wFBP) used in CT reconstructions, is the targeted weighting system. This system assigns variable importance to thin slices based on their position relative to the intended thick slice, utilizing a triangular weighting function. This approach diverges from the typical use of simple averaging and is more reflective of wFBP processes, emphasizing the core of the thick slice to enhance accuracy and reduce noise. By taking into account both the thickness and increment of slices, our algorithm achieves a level of control and flexibility in slice arrangement that is not afforded by standard averaging methods.


\bibliographystyle{IEEEtran}
\bibliography{ref.bib}

\end{document}